\documentclass{elsart}
\usepackage{amssymb}
\newcommand{\sect}[1]{\setcounter{equation}{0}\section{#1}}

\newcommand{\bfm}[1]{\mbox{\boldmath${#1}$}}

\begin{document}
\begin{frontmatter}
\title{Thermal and mechanical equilibrium among weakly
interacting systems in generalized thermostatistics framework}
\author{A.M. Scarfone}
%
\address{Istituto
Nazionale di Fisica della Materia (CNR-INFM) and Physics Department\\
Unit\'a del Politecnico di Torino, Corso Duca degli Abruzzi 24,\\
I-10129 Torino, Italy}
\date{\today}
\begin {abstract}
We consider two statistically independent systems described by the
same entropy belonging to the two-parameter family of Sharma-Mittal.
Assuming a weak interaction among the systems, allowing in this way
an exchange of heat and work, we analyze, both in the entropy
representation and in the energy representation, the evolution
toward the equilibrium. The thermodynamics evolution is controlled
by two scalar quantities identified with the temperature and the
pressure of the system. The thermodynamical stability conditions of
the equilibrium state are analyzed in both representations. Their
relationship with the concavity conditions for the entropy and with
the convexity conditions for the energy are spotlighted.
\end {abstract}
\begin{keyword}
Sharma-Mittal entropy, thermodynamical equilibrium,
thermodynamical stability. \PACS{02.50.-r, 05.20.-y,
05.90.+m}
\end{keyword}
\end{frontmatter}
\maketitle

\sect{Introduction} When two different systems, posed in
thermodynamical contact, exchange heat and work, they evolve toward
an equilibrium configuration. In the entropy representation, the
evolution toward equilibrium is controlled by the increase of
entropy, which reaches its maximum value according to the maximum
entropy principle. Differently, in the energy representation, the
evolution toward equilibrium is ruled by the decrease of energy,
which
reaches its minimum value according to the minimum energy principle.\\
As known, the formal development of the thermodynamical theory can
be equivalently carried on in both these formalisms \cite{Callen}
and many physical implications can be obtained by applying the
extremal principles. For instance, one can derive a definition of
temperature and pressure as the variables controlling the exchange
of heat and work \cite{Wannier} and obtain the thermodynamical
stability conditions (TSCs)
of the equilibrium state. \\
In this paper we will be particularly concerned with some questions
related to the approach toward the equilibrium among two weakly
interacting systems described by the same entropy belonging to the
two-parameter family of Sharma-Mittal (SM) \cite{Sharma}. Many
one-parameter entropies introduced in literature, in the framework
of the generalized statistical mechanics, belong to the SM family
and can be thus considered in a unifying scheme. Among them, we
recall the R\'{e}nyi entropy \cite{Renyi}, the Tsallis entropy
\cite{Tsallis}, the Landsberg-Vedral entropy \cite{Landsberg}, and
others \cite{Frank}. Remarkably, these entropies admit a probability
distribution function with an asymptotic power law behavior which
differs from
the exponential behavior showed by the Gibbs distribution.\\
The SM entropy, introduced initially in the information theory, has
been recently reconsidered in the framework of the generalized
thermostatistics \cite{Frank}. In \cite{Frank1} a kinetic approach
based on a nonlinear Fokker-Plank equation related to the SM entropy
has been discussed. Physical applications in the study of a weakly
interacting gas \cite{Fa} and in the context of the specific heat in
the non extensive statistical picture \cite{Lenzi} have been
reported, whilst in \cite{Masi}, it has been rediscovered on the
Kolmogorov-Nagumo
average framework \cite{Naudts}.\\
The purpose of this paper is twofold. Firstly, we study, both in the
entropy and in the energy representation, the approach toward the
equilibrium of two systems weakly interacting described by the same
entropy. It is shown that the evolution toward the equilibrium is
controlled by two scalar quantities, which can be identified with
the temperature and the pressure of the system. Alternative
definitions of temperature and pressure, in presence of a
generalized entropy, have been previously advanced in literature
\cite{Abe1,Abe2,Wada,Scarfone1}. They are based on a generalization
of the thermodynamical zero law, which is substantially different
from the dynamical approach discussed in
this work.\\
Successively, we explore the TSCs for the equilibrium state. In the
Boltzmann-Gibbs theory, TSCs are equivalent to the concavity
conditions for the entropy or to the convexity conditions for the
energy. Since the SM entropy fulfil a not linear ``composability''
rule, it is show, in accordance with the existing literature
\cite{Wada,Scarfone1,Ramshaw}, that in this case TSCs are non
equivalent to the concavity conditions for the entropy. Differently,
by assuming for the energy a linear composition, in the energy
representation TSCs are merely consequences of the convexity
conditions for the energy. In this sense, the ``composability'' rule
of the relevant physical quantities play a r\^{o}le in the
derivation of
the TSCs.\\
The plan of the paper is the following. In Section 2, we revisit the
SM entropy recalling some useful proprieties. In Section 3, we study
the approach to the equilibrium according to the maximal entropy
principle, whilst, in Section 4, we derive the TSCs in the entropy
representation. In Section 5, equilibrium and its stability are
reconsidered in the energy representation. The conclusions are
reported in Section 6.

\sect{The Sharma-Mittal entropy} Let us introduce the SM
entropy in the form
\begin{equation}
S_{q,\,{_2-r}}(p)=\ln_q\left(\sum_{i=1}^Wp_i^{2-r}\right)^{1/(r-1)}
\ ,\label{esm}
\end{equation}
(throughout this paper we take $k_{\rm B}=1$), where
\begin{equation}
\ln_q(x)=\frac{x^{1-q}-1}{1-q} \ ,\label{qlog}
\end{equation}
is the $q$-deformed logarithm and,  $q>0$ and $r<2$, are two real
parameters. In Eq. (\ref{esm}) $p\equiv\{p_i\}_{_{i=1,\cdots,W}}$ is
a discrete distribution function and $W$ denotes the
number of microstates accessible by the system.\\
For our convenience, Eq. (\ref{esm}) differs from the
definition given in Ref.
\cite{Frank} and is related to this one by $r\to2-r$. \\
Equation (\ref{esm}) includes some one-parameter entropies already
known in literature: the R\'{e}nyi entropy $S_{2-r}^{\rm
R}(p)=\ln(\sum_i p_i^{2-r})/(r-1)$ \cite{Renyi} for $q=1$, the
Tsallis entropy $S_q^{\rm T}(p)=(\sum_i p_i^q-1)/(1-q)$
\cite{Tsallis} for $r=2-q$, the Landsberg-Vedral entropy
\cite{Landsberg} $S_{2-q}^{\rm LV}=[(\sum_i
p_i^{2-q})^{-1}-1]/(1-q)$ for $r=q$, the Gaussian entropy
\cite{Frank} $S_q^{\rm G}=\ln_q[\exp(-\sum_i p_i\,\ln(p_i))]$ for
$r=1$, the escort entropy \cite{Tsallis1} $S_q^{\rm E}=[(\sum_i
p_i^{1/q})^{-q}-1]/(1-q)$ for $r=2-1/q$ and, last but not least, the
Boltzmann-Gibbs entropy $S^{\rm BG}(p)=-\sum_i p_i\,\ln(p_i)$,
recovered
in the $(q,\,r)\to(1,\,1)$ limit.\\
In the entropy representation the canonical distribution at
equilibrium can be obtained from the following variational problem
\begin{equation}
\frac{\delta}{\delta\,p_j}\left(S_{_{\{m\}}}(p)-\gamma\sum_{i=1}^Wp_i
-\beta\sum_{i=1}^Wp_i\,E_i\right)=0 \ ,\label{var}
\end{equation}
[for sake of simplicity hereinafter we introduc the notation
$\{m\}\equiv (q,\,2-r)$] where the constraints on the normalization
$\sum_ip_i=1$ and on the linear mean energy $U=\sum_ip_i\,E_i$ are
taken into account through the Lagrange multipliers $\gamma$ and
$\beta$, respectively. We observe that in Refs. \cite{Frank,Fa} the
mean energy is defined by means of ``escort'' probability
distribution. Remarkably, these two different approaches
are related according to the ``$q\to1/q$'' symmetry \cite{Frank}.\\
In a similar way, in the energy representation, the distribution at
equilibrium can be obtained from the following variational problem
\begin{equation}
\frac{\delta}{\delta\,p_j}\left(U(p)-\gamma^\prime\sum_{i=1}^Wp_i
-\beta^\prime\ln_q\left(\sum_{i=1}^Wp_{_i}^{2-r}\right)^{1/(r-1)}\right)=0
\ ,\label{var11}
\end{equation}
where now the constraints, given by the normalization and the
entropy (\ref{esm}), are taken into
account by the Lagrange multipliers $\gamma^\prime$ and $\beta^\prime$.\\
Equations (\ref{var}) and (\ref{var11}) differ only for a
redefinition of the Lagrange multipliers according to
$\beta=1/\beta^\prime$ and $\gamma=-\gamma^\prime/\beta^\prime$. As
a consequence, both the variational problems give the same
distribution
\begin{equation}
p_j={1\over \overline
Z_{_{\{m\}}}}\,\exp_r\left(-{\widetilde\beta\over2-r}\,\left(E_j-U\right)\right)
\ ,\label{dist}
\end{equation}
with the $q$-deformed exponential, the inverse function of the
$q$-deformed logarithm, given by
\begin{equation}
\exp_q(x)=[1+(1-q)\,x]_+^{\frac{1}{1-q}} \ ,
\end{equation}
and $[x]_+=$ max$(x,\,0)$ defines a cut-off condition for $r<1$,
whereas the distribution shows an asymptotic power law behavior
$p(E)\sim
E^{-1/(r-1)}$ for $r>1$.\\
The quantity $\widetilde\beta$ is given by
\begin{equation}
\widetilde\beta={\beta\over\left(\overline
Z_{_{\{m\}}}\right)^{1-q}} ={\beta\over1+(1-q)\,S_{_{\{m\}}}} \
,\label{bb}
\end{equation}
and the normalization function $\overline Z_{_{\{m\}}}$, defined in
\begin{equation}
\overline Z_{_{\{m\}}}=\left(\sum_{i=1}^Wp_i^{2-r}\right)^{1/(r-1)}
\ ,\label{z}
\end{equation}
is a function of the Lagrange multipliers $\gamma$ and $\beta$
\begin{equation}
\left(\overline
Z_{_{\{m\}}}\right)^{1-q}={r-1\over2-r}\Big(\gamma+\beta\,U\Big) \ .
\end{equation}
The function $\overline Z_{_{\{m\}}}$ is related to the canonical
partition function $Z_{_{\{m\}}}$ in
\begin{equation}
\ln_q\left(Z_{_{\{m\}}}\right)=\ln_q\left(\overline
Z_{_{\{m\}}}\right)-\beta\,U \ ,
\end{equation}
so that, from the definition (\ref{esm}) we obtain
\begin{equation}
S_{_{\{m\}}}=\ln_q\left(Z_{_{\{m\}}}\right)+\beta\,U \ .\label{leg}
\end{equation}
This relationship between the entropy and the partition function,
through the introduction of a suitable deformed logarithm, is
recurrent in different generalized formulations of the statistical
mechanics (in addition to the classical Boltzmann-Gibbs theory)
\cite{Tsallis2,Scarfone4,Scarfone5}.\\
Because $\exp_q(x)$ is a monotonic and increasing function, from Eq.
(\ref{dist}) it could appear that the most probable state
corresponds to the fundamental energy level. On the other hand, let
us introduce the multiplicity $\Omega({\mathcal E}_j,\,V)$ of a
macrostate with energy ${\mathcal E}_j$. It depends on the volume
$V$ of the system and represents the number of possible microstates
with the same energy ${\mathcal E}_j$. By taking into account that
all the probabilities $p_i$ of the microstates belonging to the same
macrostate, labeled by the energy ${\mathcal E}_j$, have the same
value, we can introduce the relevant probability $P({\mathcal
E}_j,\,V)$ of a macrostate as
\begin{equation}
P({\mathcal E}_j,\,V)={\Omega({\mathcal E}_j,\,V)\over \overline
Z_{_{\{m\}}}}\,\exp_r\left(-\frac{\widetilde\beta}{2-r}\,\left({\mathcal
E}_j-U\right)\right) \ .\label{distr}
\end{equation}
Therefore, the most probable state, which maximize the relevant
probability $P({\mathcal E}_j,\,V)$, is given by the competition
among $p_j$, which is a monotonic decreasing function with respect
to the energy and $\Omega({\mathcal E}_j,\,V)$ which is typically
a monotonic increasing function.\\
In the microcanonical picture, since all the microstates have the
same energy $U$, the relevant probability is given by $P(U,\,V)=1$
because $P({\mathcal E}_j,\,V)=0$ for ${\mathcal E}_j\not=U$. In
this case $\Omega(U,\,V)=W$, Eq. (\ref{dist}) reduces to the uniform
distribution with $p_i=1/W$ and entropy (\ref{esm}) assumes the
expression
\begin{equation}
S_{_{\{m\}}}\left(U,\,V\right)=\ln_q\Big(W\Big) \ .\label{bg}
\end{equation}
Equation (\ref{bg}) mimics the Boltzmann formula for the entropy,
recovered in the $q\to1$
limit.\\
In the following we derive some proprieties of entropy
$S_{_{\{m\}}}(U,\,V)$ useful in the next sections.\\
Firstly, from Eq. (\ref{var}), using Eq. (\ref{z}), it follows
\begin{equation}
{2-r\over r-1}\,\left(\overline
Z_{_{\{m\}}}\right)^{r+q-2}\,p_j^{1-r}=\gamma+\beta\,E_j \ ,
\end{equation}
and, after deriving Eq. (\ref{esm}), we obtain the relation
\begin{equation}
dS_{_{\{m\}}}={2-r\over r-1}\left(\overline
Z_{_{\{m\}}}\right)^{r+q-2}\sum_{j=1}^Wp_j^{1-r}\,dp_j=\sum_{j=1}^W\left(\gamma+\beta\,E_j\right)\,dp_i
\ .\label{eu}
\end{equation}
By taking into account that $\sum_idp_{_i}=0$, as it follows from
the normalization on $p_{_i}$, and recalling that
$dU=\sum_idE_i\,p_i+\sum_iE_i\,dp_i\equiv\delta {\mathcal L}+\delta
{\mathcal Q}$ (first law), under the ``no work'' condition $\delta
{\mathcal L}\equiv\sum_idE_i\,p_i=0$, we obtain the fundamental
thermodynamics relation
\begin{equation}
\left(\frac{\partial\,S_{_{\{m\}}}}{\partial\,U}\right)_{\rm
V}=\beta \ .\label{eu1}
\end{equation}
Moreover, from the definition (\ref{bb}) we have
\begin{eqnarray}
\hspace{-10mm}\left(\frac{\partial\,\widetilde\beta}{\partial\,U}\right)_{\rm
V}={1\over1+(1-q)\,S_{_{\{m\}}}}\left[\frac{\partial^2
S_{_{\{m\}}}}{\partial
U^2}-{1-q\over1+(1-q)\,S_{_{\{m\}}}}\,\left(\frac{\partial
S_{_{\{m\}}}}{\partial U}\right)^2\right]_{\rm V} \ ,\label{eeu2}
\end{eqnarray}
which, for a stable equilibrium configuration, implies
\begin{equation}
\left(\frac{\partial\,\widetilde\beta}{\partial\,U}\right)_{\rm V}<0
\ ,\label{eu2}
\end{equation}
(a sketch of this statement will be given in Section 4
[cfr. Eq. (\ref{prima})]).\\
Secondly, we recall the ``composability'' rule of entropy
$S_{_{\{m\}}}(U,\,V)$ for two statistically independent systems A
and B, in the sense of $p^{{\rm A}\cup{\rm B}}_{ij}=p^{^{\rm
A}}_i\cdot p^{{\rm B}}_j$. From the definition (\ref{esm}) it
follows that
\begin{equation}
S_{_{\{m\}}}^{{\rm A}\cup{\rm B}}=S_{_{\{m\}}}^{\rm
A}+S_{_{\{m\}}}^{\rm B}+(1-q)\,S_{_{\{m\}}}^{\rm
A}\,S_{_{\{m\}}}^{\rm B} \ ,\label{add}
\end{equation}
where the ``super-additivity'' ($q<1$) and the ``sub-additivity''
($q>1$) behaviors are controlled only by the parameter $q$. Linear
composability is recovered for $q=1$, i.e. for the R\'{e}nyi
family.\\Finally, we discuss the concavity conditions of entropy and
the convexity conditions of energy. As it is well known, the
concavity conditions for the given problem follow from the analysis
of the sign of the eigenvalues of the Hessian matrix associated to
the function $S_{_{\{m\}}}(U,\,V)$. In particular, by requiring that
the following quadratic form
\begin{equation}
\phi({\bfm y})=\frac{\partial^2S_{_{\{m\}}}}{\partial U^2}\,y_{\rm
U}^2+2\,\frac{\partial^2S_{_{\{m\}}}}{\partial U\,\partial
V}\,y_{\rm U} \,y_{\rm V}+\frac{\partial^2S_{_{\{m\}}}}{\partial
V^2}\,y_{\rm V}^2 \label{quad}
\end{equation}
be negative definite for any arbitrary vector ${\bfm y}\equiv(y_{\rm
U},\,y_{\rm V})$, we obtain the relations
\begin{equation}
\frac{\partial^2S_{_{\{m\}}}}{\partial U^2}<0 \ ,\hspace{10mm}
\frac{\partial^2S_{_{\{m\}}}}{\partial
U^2}\,\frac{\partial^2S_{_{\{m\}}}}{\partial V^2}-
\left(\frac{\partial^2S_{_{\{m\}}}}{\partial U\,\partial
V}\right)^2>0 \ ,\label{cc}
\end{equation}
stating the concavity conditions of the SM entropy (remark that Eqs.
(\ref{cc}) imply the further relation
$\partial^2S_{_{\{m\}}}/\partial V^2<0$).\\
In a similar way, the energy $U(S_{_{\{m\}}},\,V)$ is a convex
function with respect to $S_{_{\{m\}}}$ and $V$ if the following
quadratic form
\begin{equation}
\psi({\bfm y})=\frac{\partial^2U}{\partial S_{_{\{m\}}}^2}\,y_{\rm
S}^2+2\,\frac{\partial^2U}{\partial S_{_{\{m\}}}\,\partial
V}\,y_{\rm S} \,y_{\rm V}+\frac{\partial^2U}{\partial V^2}\,y_{\rm
V}^2 \label{quad1}
\end{equation}
is positive definite for any arbitrary vector ${\bfm y}\equiv(y_{\rm
S},\,y_{\rm V})$. Easily, we obtain the relations
\begin{equation}
\frac{\partial^2U}{\partial S_{_{\{m\}}}^2}>0 \
,\hspace{10mm}\frac{\partial^2U}{\partial
S_{_{\{m\}}}^2}\,\frac{\partial^2U}{\partial V^2}-
\left(\frac{\partial^2U}{\partial S_{_{\{m\}}}\,\partial
V}\right)^2>0 \ ,\label{ccu}
\end{equation}
which state the convexity conditions for the energy.
\sect{Thermal and mechanical equilibrium}

We consider an initial situation where two isolated systems A and B,
with mean energies $U^{\rm A}$ and $U^{\rm B}$ and volumes $V^{\rm
A}$ and $V^{\rm B}$, respectively, are described by the same
entropy. Exchange of heat (energy) and work (volume), which may take
place between the
systems, are initially prohibited.\\
Latter, some constraints are relaxed allowing, in this way, to
establish weak interactions among them.\\ The whole system A$\cup$B
is now subjected to new constraints given by the total energy
$U^{{\rm A}\cup{\rm B}}=U^{\rm A}+U^{\rm B}$ and the total volume
$V^{{\rm A}\cup{\rm B}}=V^{\rm A}+V^{\rm B}$ which we assume to be
conserved in time. This means that we are neglecting the interaction
among the two systems. In fact, in the limit of zero interaction the
energies and volumes are strictly additive, however, a small
interaction between the parts is required to enable some exchange of
heat and work among them. In the same way, we can pose
\begin{equation}
p_{ij}^{{\rm A}\cup{\rm B}}=p_i^{\rm A}\,p_j^{\rm
B}\,(1+\delta p_{ij}) \ ,
\end{equation}
where $\delta p_{ij}$ takes into account the correlations between
the systems. Under the hypothesis of very weak interaction it is
reasonable to neglect this term, i.e., we assume that the
statistical independence among the systems is preserved in time.\\
When the systems are posed in thermodynamical contact, the entropy
is not at its maximum value due to the new constraints. The system
will evolve toward a new equilibrium increasing its entropy, $\delta
S_{_{\{m\}}}^{{\rm A}\cup{\rm B}}>0$, until
reaches its extreme limit.\\
By evaluating the variation of $S_{_{\{m\}}}^{{\rm A}\cup{\rm B}}$,
up to the first order in $\delta U$ and $\delta V$, from Eq.
(\ref{add}) we obtain
\begin{eqnarray}
\nonumber & &\!\!\!\delta S_{_{\{m\}}}^{{\rm A}\cup{\rm
B}}=\left[1+(1-q)\,S_{_{\{m\}}}^{\rm
B}\right]\left[\left(\frac{\partial S_{_{\{m\}}}^{\rm A}}{\partial
U^{\rm A}}\right)_{\rm U}\delta U+\left(\frac{\partial
S_{_{\{m\}}}^{\rm A}}{\partial V^{\rm
A}}\right)_{\rm V}\delta V\right]\\
\nonumber & &\;\;\;\;\;\;\;\;\;-\left[1+(1-q)\,S_{_{\{m\}}}^{\rm
A}\right]\left[\left(\frac{\partial S_{_{\{m\}}}^{\rm B}}{\partial
U^{\rm B}}\right)_{\rm U}\delta
U+\left(\frac{\partial\,S_{_{\{m\}}}^{\rm B}}{\partial\,V^{\rm
B}}\right)_{\rm V}\,\delta V\right]\\
\nonumber & &\;\;\;\;\;\;\;\;\;=\left[1+(1-q)\,S_{_{\{m\}}}^{\rm
A}\right]\left[1+(q-1)\,S_{_{\{m\}}}^{\rm B}\right]\\
\nonumber &
&\!\!\!\!\!\times\left\{\left[{1\over1+(1-q)\,S_{_{\{m\}}}^{\rm
A}}\left(\frac{\partial S_{_{\{m\}}}^{\rm A}}{\partial U^{\rm
A}}\right)_{\rm V}-{1\over1+(1-q)\,S_{_{\{m\}}}^{\rm
B}}\left(\frac{\partial S_{_{\{m\}}}^{\rm B}}{\partial U^{\rm
B}}\right)_{\rm V}\right] \delta U\right.\\
\nonumber &
&\!\!\!\!\!+\left.\left[{1\over1+(1-q)\,S_{_{\{m\}}}^{\rm
A}}\left(\frac{\partial S_{_{\{m\}}}^{\rm A}}{\partial V^{\rm
A}}\right)_{\rm U}-{1\over1+(1-q)\,S_{_{\{m\}}}^{\rm
B}}\left(\frac{\partial S_{_{\{m\}}}^{\rm B}}{\partial
V^{\rm B}}\right)_{\rm U}\right]\delta V\right\}>0 \ ,\\
& &\label{var1}
\end{eqnarray}
where, we pose $\delta U^{\rm A}=-\delta U^{\rm B}\equiv \delta U$
and $\delta V^{\rm A}=-\delta V^{\rm B}\equiv \delta V$, according
to the conservation of $U^{{\rm A}\cup{\rm B}}$ and $V^{{\rm
A}\cup{\rm B}}$.\\ Assuming firstly $\delta V=0$, from Eq.
(\ref{var1}) it follows
\begin{equation}
\left[{\beta^{\rm A}\over1+(1-q)\,S_{_{\{m\}}}^{\rm A}}-{\beta^{\rm
B}\over1+(1-q)\,S_{_{\{m\}}}^{\rm B}}\right]\,\delta U>0 \
,\label{bbeta}
\end{equation}
since $1+(1-q)\,S_{_{\{m\}}}>0$. Equation (\ref{bbeta}) can be
written in
\begin{equation}
\left(\widetilde\beta^{\rm A}-\widetilde\beta^{\rm B}\right)\,\delta
U>0 \ ,\label{beta}
\end{equation}
which implies that ${\rm sgn}(\delta U)={\rm
sgn}(\widetilde\beta^{\rm A}-\widetilde\beta^{\rm B})$\footnote{The
sign function sgn$(x)$ is defined in sgn$(x)=+1$ for $x>0$ and
sgn$(x)=-1$ for $x<0$.}. This means that energy flows always from
the system with smaller $\widetilde\beta$ to the system with larger
$\widetilde\beta$. Such a process goes on until the equilibrium,
stated by the equality $\widetilde\beta^{\rm A}=\widetilde\beta^{\rm
B}$, is
reached.\\
The main facts of $\widetilde \beta$ reflect the same physical
proprieties of $\beta=1/T$ of the standard thermodynamics
\cite{Wannier} and can be summarized in the following
points:\\
a) Two systems which cannot exchange energy have
in general
different $\widetilde\beta$'s.\\
b) When two weakly interacting systems exchange energy, their
respective values of $\widetilde\beta$ become equal when equilibrium
is reached.\\
c) Between two weakly interacting systems the energy flows always
from the system with the smaller $\widetilde\beta$ to the system
with the larger
$\widetilde\beta$.\\
d) The mean energy of a system, in a stable equilibrium
configuration, increases monotonically as $\widetilde\beta$
decreases, according to Eq.
(\ref{eu2}).\\
The parameter $\widetilde\beta$ is a variable controlling the
exchange of energy among the systems and can be identified with the
temperature according to
\begin{equation}
T={1\over \widetilde\beta} \ .\label{temp}
\end{equation}
Equations (\ref{bb}), (\ref{eu1}) and (\ref{temp}) establish a
relationships between temperature and entropy which differs from the
standard one $\beta=1/T=(\partial S/\partial U)_{\rm V}$. In other
words, the inverse of the temperature differs form the Lagrange
multiplier $\beta$ associated to the mean energy $U$. The standard
relationship is recovered only in the $q\to1$ limit, where the SM
entropy reduces to the R\'{e}nyi entropy. For the Tsallis' case,
with $r=2-q$, Eq. (\ref{temp}) coincides with the definition of
temperature obtained in Ref. \cite{Abe1} and derived by means of
separability constant between the thermal equilibrium of two
systems.\\ In the microcanonical picture Eq. (\ref{temp}) coincides
with to the standard definition of the Boltzmann temperature. In
fact, in this case Eq. (\ref{dist}) collapses to the uniform
distribution $p_i=1/W$ and, accounting for Eq. (\ref{bg}), we obtain
\begin{equation}
{1\over T}=\Bigg(\frac{\partial}{\partial U} \ln W\Bigg)_{\rm V} \ .
\end{equation}
This fact agrees with already known results obtained by using the
Tsallis' entropy \cite{Toral} and the
Sharma-Taneja-Mittal entropy \cite{Scarfone1,Scarfone2}.\\
Similar arguments can be applied to obtain a definition of pressure.
In fact, by posing $\delta U=0$, from Eq. (\ref{var1}) we obtain
\begin{equation}
\hspace{-4mm}\left[{1\over1+(1-q)\,S_{_{\{m\}}}^{\rm
A}}\left(\frac{\partial S_{_{\{m\}}}^{\rm A}}{\partial V^{\rm
A}}\right)_{\rm U}-{1\over1+(1-q)\,S_{_{\{m\}}}^{\rm
B}}\left(\frac{\partial S_{_{\{m\}}}^{\rm B}}{\partial V^{\rm
B}}\right)_{\rm U}\right]\delta V>0 \ ,\label{pr}
\end{equation}
which advances the following definition
\begin{equation}
P={T\over1+(1-q)\,S_{_{\{m\}}}}\left(\frac{\partial
S_{_{\{m\}}}}{\partial V}\right)_{\rm U} \ .\label{pres}
\end{equation}
Recalling that $\delta U=0$ implies thermal equilibrium, i.e.
$\widetilde\beta^{\rm A}=\widetilde\beta^{\rm B}\equiv\beta>0$, Eq.
(\ref{pr}) can be rewritten in
\begin{equation}
\left(P^{\rm A}-P^{\rm B}\right)\delta V\geq0 \ ,\label{pr1}
\end{equation}
so that sgn($\delta V)={\rm sgn}(P^{\rm A}-P^{\rm B})$, which means
that the system with greater pressure increases
its volume, whilst the system with lowest pressure reduces its volume.\\
It is worth to observe that, by taking into account Eq. (\ref{eu1})
and the relation $(\partial U/\partial S_{_{\{m\}}})_{\rm
V}\,(\partial S_{_{\{m\}}}/\partial V)_{\rm U}=-(\partial U/\partial
V)_{\rm S}$, Eq. (\ref{pres}) can be written in
\begin{equation}
P=-\left(\frac{\partial U}{\partial V}\right)_{\rm S} \ ,\label{pr2}
\end{equation}
which coincides with the definition of pressure given in the
standard thermostatistics. What is different, as stated by Eq.
(\ref{pres}), is the relationships between the pressure $P$ and the
entropy $S_{_{\{m\}}}(U,\,V)$.

\sect{Thermodynamical stability}

The thermodynamical stability conditions for the entropy
$S_{_{\{m\}}}(U,\,V)$ can be obtained by analyzing the sign of the
entropy changes produced by perturbing the system away from the
equilibrium. To begin with, we expand the variation of the entropy
$\delta S_{_{\{m\}}}^{{\rm A}\cup{\rm B}}$ up to the second order in
$\delta U$ and $\delta V$. By recalling that at the equilibrium the
first order terms vanish, we obtain
\begin{eqnarray}
\nonumber
&&\hspace{-7mm}{1\over2}\left\{\left[1+(1-q)\,S_{_{\{m\}}}^{\rm
B}\right]\left[\frac{\partial^2 S_{_{\{m\}}}^{\rm A}}{\left(\partial
U^{\rm A}\right)^2}-\frac{1-q}{1+(1-q)\,S^{\rm
A}_{_{\{m\}}}}\left(\frac{\partial S_{_{\{m\}}}^{\rm A}}{\partial
U^{\rm
A}}\right)^2\right]\right.\\
\nonumber& &\hspace{-5mm}\left.+\left[1+(1-q)\,S_{_{\{m\}}}^{\rm
A}\right]\left[\frac{\partial^2 S_{_{\{m\}}}^{\rm B}}{\left(\partial
U^{\rm B}\right)^2}-\frac{1-q}{1+(1-q)\,S^{\rm
B}_{_{\{m\}}}}\left(\frac{\partial S_{_{\{m\}}}^{\rm B}}{\partial
U^{\rm B}}\right)^2\right]\right\}\,(\delta
U)^2\\
\nonumber&&\hspace{-10mm}+{1\over2}\left\{\left[1+(1-q)\,S_{_{\{m\}}}^{\rm
B}\right]\left[\frac{\partial^2 S_{_{\{m\}}}^{\rm A}}{\left(\partial
V^{\rm A}\right)^2}-\frac{1-q}{1+(1-q)\,S^{\rm
A}_{_{\{m\}}}}\left(\frac{\partial S_{_{\{m\}}}^{\rm A}}{\partial
V^{\rm
A}}\right)^2\right]\right.\\
\nonumber& &\hspace{-5mm}\left.+\left[1+(1-q)\,S_{_{\{m\}}}^{\rm
A}\right]\left[\frac{\partial^2 S_{_{\{m\}}}^{\rm B}}{\left(\partial
V^{\rm B}\right)^2}-\frac{1-q}{1+(1-q)\,S^{\rm
B}_{_{\{m\}}}}\left(\frac{\partial S_{_{\{m\}}}^{\rm B}}{\partial
V^{\rm B}}\right)^2\right]\right\}\,(\delta
V)^2\\
\nonumber & &\hspace{-7mm}+\left\{\left[1+(1-q)\,S_{_{\{m\}}}^{\rm
B}\right]\left[\frac{\partial^2 S_{_{\{m\}}}^{\rm A}}{\partial
U^{\rm A}\partial V^{\rm A}}-\frac{1-q}{1+(1-q)\,S^{\rm
A}_{_{\{m\}}}}\frac{\partial S_{_{\{m\}}}^{\rm A}}{\partial U^{\rm
A}}\frac{\partial S_{_{\{m\}}}^{\rm A}}{\partial V^{\rm
A}}\right]\right.\\
\nonumber& &\hspace{-10mm}\left.+\left[1+(1-q)\,S_{_{\{m\}}}^{\rm
A}\right]\left[\frac{\partial^2 S_{_{\{m\}}}^{\rm B}}{\partial
U^{\rm B}\partial V^{\rm B}}-\frac{1-q}{1+(1-q)\,S^{\rm
B}_{_{\{m\}}}}\frac{\partial S_{_{\{m\}}}^{\rm B}}{\partial U^{\rm
B}}\frac{\partial S_{_{\{m\}}}^{\rm B}}{\partial V^{\rm
B}}\right]\right\}\,\delta U\delta V<0 ,\\ &
&\hspace{-15mm}\label{var2}
\end{eqnarray}
and introducing the notation
\begin{equation}
{\mathcal S}_{\rm XY}=\frac{\partial^2 S_{_{\{m\}}}}{\partial
X\,\partial Y}-\frac{1-q}{1+(1-q)\,S_{_{\{m\}}}}\,\frac{\partial
S_{_{\{m\}}}}{\partial X}\,\,\frac{\partial S_{_{\{m\}}}}{\partial
Y} \ ,
\end{equation}
where $X$ and $Y$ stand for $U$ or $V$, Eq. (\ref{var2}) can be
written as
\begin{eqnarray}
\nonumber & &{1\over2}\,\left[1+(1-q)\,S_{_{\{m\}}}^{\rm
B}\right]\left[{\mathcal S}^{\rm A}_{UU}\,\left(\delta
U\right)^2+{\mathcal S}^{\rm A}_{UV}\,\delta U\,\delta V+{\mathcal
S}^{\rm
A}_{VV}\,\left(\delta V\right)^2\right]\\
& +&{1\over2}\,\left[1+(1-q)\,S_{_{\{m\}}}^{\rm
A}\right]\left[{\mathcal S}^{\rm B}_{UU}\,\left(\delta
U\right)^2+{\mathcal S}^{\rm B}_{UV}\,\delta U\,\delta V+{\mathcal
S}^{\rm B}_{VV}\,\left(\delta V\right)^2\right]<0 \ .\label{var3}
\end{eqnarray}
This equation is fulfilled if the following inequalities
\begin{equation}
{\mathcal S}_{UU}<0 \ ,\hspace{10mm}{\mathcal
S}_{UU}\,{\mathcal S}_{VV}-\left({\mathcal
S}_{UV}\right)^2>0 \ ,\label{1}
\end{equation}
are separately satisfied by both systems. [A further
relation ${\mathcal S}_{_{VV}}<0$ follows from Eqs. (\ref{1})].\\
Explicitly, we have
\begin{eqnarray}
&&\frac{\partial^2S_{_{\{m\}}}}{\partial
U^2}<{1-q\over1+(1-q)\,S_{_{\{m\}}}}\,\left(\frac{\partial
S_{_{\{m\}}}}{\partial U}\right)^2 \ ,\label{prima}\\
&&\nonumber\\ &&\frac{\partial^2S_{_{\{m\}}}}{\partial
U^2}\,\frac{\partial^2S_{_{\{m\}}}}{\partial
V^2}-\left(\frac{\partial^2S_{_{\{m\}}}}{\partial U\,\partial
V}\right)^2>{1-q\over1+(1-q)\,S_{_{\{m\}}}}\,{\mathcal B}_{_{\{m\}}}
\ ,\label{seconda}
\end{eqnarray}
where
\begin{eqnarray}
\nonumber {\mathcal B}_{_{\{m\}}}&=&\left(\frac{\partial^2
S_{_{\{m\}}}}{\partial U^2}\right)^{-1}\left\{\left(\frac{\partial^2
S_{_{\{m\}}}}{\partial U^2}\frac{\partial S_{_{\{m\}}}}{\partial
V}-\frac{\partial^2 S_{_{\{m\}}}}{\partial U\,\partial
V}\frac{\partial S_{_{\{m\}}}}{\partial
U}\right)^2\right.\\
&+&\left.\left(\frac{\partial S_{_{\{m\}}}}{\partial
U}\right)^2\left[\frac{\partial^2S_{_{\{m\}}}}{\partial
U^2}\,\frac{\partial^2S_{_{\{m\}}}}{\partial V^2}-
\left(\frac{\partial^2 S_{_{\{m\}}}}{\partial U\,\partial
V}\right)^2\right]\right\} \ ,\label{cb}
\end{eqnarray}
is a negative definite quantity for a concave entropy.\\ Equations
(\ref{prima})-(\ref{seconda}) are the thermodynamical stability
conditions for the entropies belonging to the SM family. They reduce
to the concavity
conditions (\ref{cc}) in the $q\to1$ limit.\\
We observe that for ``super-additive'' and ``additive'' systems,
with $q\leq1$, Eqs. (\ref{prima})-(\ref{seconda}) are satisfied if
the concavity conditions (\ref{cc}) holds. Moreover, when $q<1$ the
equilibrium configuration is stable also if the entropy shows a
small convexity. In this sense the ``super-additive'' systems
exhibit a kind of ``super-stability''. We observe that Eq.
(\ref{prima}) implies Eq. (\ref{eu2}) so that, if the TSCs are
fulfilled, among two bodies in thermal contact heat always flows
from hot body to cold body,
freely from the concavity arguments on the entropy.\\
Differently, for ``sub-additive'' systems with $q>1$, the concavity
conditions are not enough to guarantee the thermodynamical stability
of the equilibrium configuration. In this case, entropies with a not
very pronounced concavity, can still violate Eqs.
(\ref{prima})-(\ref{seconda}) as well as Eq. (\ref{eu2}). Thus, we
can state a kind of ``sub-stability'' for ``sub-additive'' systems
according to Eqs. (\ref{prima})-(\ref{seconda}).

\sect{Equilibrium and stability in the energy
representation}

The study of the approach in the direction of the equilibrium and
the analysis of its stability in energy representation require only
a straightforward
transcription of language.\\
Let us consider two isolated systems A and B initially at
equilibrium, constrained by their respective entropies
$S_{_{\{m\}}}^{\rm A},\,S_{_{\{m\}}}^{\rm B}$ and volumes $V^{\rm
A},\,V^{\rm B}$. When certain constraints are removed, a weak
interaction among the systems starts on, giving origin to an
exchange of heat (entropy) and work (volume). We assume the total
entropy $S^{{\rm A}\cup{\rm B}}_{_{\{m\}}}=S^{\rm
A}_{_{\{m\}}}+S^{\rm B}_{_{\{m\}}}+(1-q)\,S^{\rm
A}_{_{\{m\}}}\,S^{\rm B}_{_{\{m\}}}$ and the total volume $V^{{\rm
A}\cup{\rm B}}=V^{\rm A}+V^{\rm B}$ constants in time.\\ According
to the minimum energy principle the system will evolve toward a new
equilibrium with lower energy. By evaluating, up the first order in
$\delta S_{_{\{m\}}}$ and $\delta V$, the changing in the energy
$U^{{\rm A}\cup{\rm B}}$, we obtain
\begin{eqnarray}
\nonumber \delta U^{{\rm A}\cup{\rm B}}&=&\left[\left(\frac{\partial
U^{\rm A}}{\partial S_{_{\{m\}}}^{\rm A}}\right)_{\rm V}\delta
S^{\rm A}_{_{\{m\}}}+\left(\frac{\partial U^{\rm B}}{\partial
S_{_{\{m\}}}^{\rm B}}\right)_{\rm V}\delta S^{\rm
B}_{_{\{m\}}}\right]\\&+&\left[\left(\frac{\partial U^{\rm
A}}{\partial V^{\rm A}}\right)_{\rm S}\delta V^{\rm
A}+\left(\frac{\partial U^{\rm B}}{\partial V^{\rm B}}\right)_{\rm
S}\delta V^{\rm B}\right]<0 \ ,\label{varu}
\end{eqnarray}
and by posing $\delta S^{\rm
A}_{_{\{m\}}}/[1+(1-q)\,S_{_{\{m\}}}^{\rm A}]=-\delta S^{\rm
B}_{_{\{m\}}}/[1+(1-q)\,S_{_{\{m\}}}^{\rm B}]\equiv \delta \Sigma$
and $\delta V^{\rm A}=-\delta V^{\rm B}\equiv \delta V$, Eq.
(\ref{varu}) becomes
\begin{eqnarray}
\nonumber\hspace{-8mm} \delta U^{{\rm A}\cup{\rm
B}}&=&\left[\Big[1+(1-q)\,S_{_{\{m\}}}^{\rm
A}\Big]\left(\frac{\partial U^{\rm A}}{\partial S_{_{\{m\}}}^{\rm
A}}\right)_{\rm V}-\Big[1+(1-q)\,S_{_{\{m\}}}^{\rm
A}\Big]\left(\frac{\partial U^{\rm B}}{\partial S_{_{\{m\}}}^{\rm
B}}\right)_{\rm V}\right]\,\delta
\Sigma\\&+&\left[\left(\frac{\partial U^{\rm A}}{\partial V^{\rm
A}}\right)_{\rm S}-\left(\frac{\partial U^{\rm B}}{\partial V^{\rm
B}}\right)_{\rm S}\right]\delta V \ .\label{var5}
\end{eqnarray}
Assuming firstly $\delta V=0$ and by taking into account Eqs.
(\ref{bb}) and (\ref{eu1}), from Eq. (\ref{var5}) we obtain
\begin{equation}
\left({1\over\widetilde\beta^{\rm A}}-{1\over\widetilde\beta^{\rm
B}}\right)\,\delta \Sigma\leq0 \ ,\label{beta1}
\end{equation}
\vspace{-4mm}which is equivalent to Eq. (\ref{beta}). We remark only
that Eq. (\ref{beta1}) is a relation in $S_{_{\{m\}}}$ and $V$ which
is more conveniently assumed when the total energy $U^{{\rm
A}\cup{\rm B}}$ is known, whereas Eq. (\ref{beta}) is a relation in
$U$ and $V$ which is more conveniently assumed when the total
entropy $S_{_{\{m\}}}^{{\rm A}\cup{\rm B}}$ is known.\\
In a similar way, by posing $\delta \Sigma=0$, from Eq. (\ref{var5})
we obtain
\begin{equation}
\left[\left(\frac{\partial\,U^{\rm A}}{\partial\,V^{\rm
A}}\right)_{\rm S}-\left(\frac{\partial\,U^{\rm
B}}{\partial\,V^{\rm B}}\right)_{\rm S}\right]\,\delta
V\leq0 \ ,
\end{equation}
which coincides with Eq. (\ref{pr1}) according to the definition (\ref{pr2}).\\
Finally, in order to obtain the thermodynamical stability conditions
in the energy representation we proceed by expanding the variation
of the energy $\delta U^{{\rm A}\cup{\rm B}}$ around the
equilibrium, up to the second order in $\delta \Sigma$ and $\delta
V$. We obtain
\begin{eqnarray}
\nonumber &{1\over2}&\left\{\frac{\partial^2 U^{\rm
A}}{\left(\partial S_{_{\{m\}}}^{\rm
A}\right)^2}\,\left[1+(1-q)\,S^{\rm
A}_{_{\{m\}}}\right]^2+\frac{\partial^2 U^{\rm B}}{\left(\partial
S_{_{\{m\}}}^{\rm B}\right)^2}\,\left[1+(1-q)\,S^{\rm
B}_{_{\{m\}}}\right]^2\right\}\,\left(\delta
\Sigma\right)^2\\
\nonumber &+&\left\{\frac{\partial^2 U^{\rm A}}{\partial
S_{_{\{m\}}}^{\rm A}\partial V^{\rm A}}\,\left[1+(1-q)\,S^{\rm
A}_{_{\{m\}}}\right]+\frac{\partial^2 U^{\rm B}}{\partial
S_{_{\{m\}}}^{\rm B}\,\partial V^{\rm B}}\,\left[1+(1-q)\,S^{\rm
B}_{_{\{m\}}}\right]\right\}\,\delta \Sigma\delta V\\
&+&{1\over2}\left[\frac{\partial^2 U^{\rm A}}{\left(V^{\rm
A}\right)^2}+\frac{\partial^2 U^{\rm A}}{\left(V^{\rm
A}\right)^2}\right]\, \left(\delta V\right)^2>0 \ ,\label{var4}
\end{eqnarray}
and recalling that $1+(1-q)\,S_{_{\{m\}}}>0$, from Eq. (\ref{var4})
it follows
\begin{equation}
\frac{\partial^2U}{\partial S_{_{\{m\}}}^2}>0 \ ,\hspace{20mm}
\frac{\partial^2U}{\partial
S_{_{\{m\}}}^2}\,\frac{\partial^2U}{\partial
V^2}-\left(\frac{\partial^2U}{\partial S_{_{\{m\}}}\,\partial
V}\right)^2>0 \ ,\label{tse}
\end{equation}
that coincides with the convexity conditions for the energy
(\ref{ccu}). This result was expected and shows us that when the
``composability rule'' of a thermodynamical quantity is linear, like
the energy in the present case, the structures of the TSCs are
equivalent to concavity (convexity) arguments of the same quantity,
freely from the ``composability'' proprieties of the other
thermodynamical quantities (for a discussion of a thermostatistics
theory based on non linear additive energies see, for instance, Ref.
\cite{Wang}).

\sect{Conclusions}

In this work we have studied the thermodynamical equilibrium and its
stability among two systems weakly interacting and described by the
same entropy belonging to the family of Sharma-Mittal. We have
derived a definition of temperature and pressure, controlling the
exchange of heat and work between the two systems. It has been shown
that temperature and pressure, obtained from dynamical arguments,
coincides with the ones already known in literature and derived from
statical considerations. We have inquired on the TSCs both in the
entropy and in the energy representation. It is shown that, due to
the nonlinear ``composability'' rule of entropy, the concavity
conditions alone are not necessary nor sufficient conditions for the
stability of the equilibrium. In particular, when the system is
sub-additive the concavity conditions do not imply the stability
whereas, when the system is super-additive, the concavity conditions
imply the stability of the equilibrium configuration. A different
situation is obtained in the energy representation where, by
assuming a linear ``composability'', the TSCs imply the convexity
conditions for the energy. This shows that, although the two
representations are equivalent, the analysis of some thermodynamical
proprieties, like for instance the TSCs, could be performed more
easily in one than in the other representation.


\end{document}